\documentclass[aps,prc,letterpaper,11pt,twoside,tightenlines,nofootinbib,showpacs,preprint]{revtex4}
\usepackage{graphicx}
\usepackage[sort&compress]{natbib}
\usepackage{subfigure}
\usepackage{amsmath}
\usepackage{amsfonts}
\usepackage{cancel}
\usepackage{lmodern,dsfont}
\usepackage{amssymb}
\begin{document}
\arraycolsep1.5pt
\newcommand{\Ima}{\textrm{Im}}
\newcommand{\Rea}{\textrm{Re}}
\newcommand{\mev}{\textrm{ MeV}}
\newcommand{\be}{\begin{equation}}
\newcommand{\ee}{\end{equation}}
\newcommand{\bea}{\begin{eqnarray}}
\newcommand{\eea}{\end{eqnarray}}
\newcommand{\bef}{\begin{figure}}
\newcommand{\eef}{\end{figure}}
\newcommand{\bce}{\begin{center}}
\newcommand{\ece}{\end{center}}
\newcommand{\ba}{\begin{eqnarray}}
\newcommand{\ea}{\end{eqnarray}}
\newcommand{\gev}{\textrm{ GeV}}
\newcommand{\nn}{{\nonumber}}
\newcommand{\dtres}{d^{\hspace{0.1mm} 3}\hspace{-0.5mm}}
\newcommand{\rts}{ \sqrt s}
\newcommand{\non}{\nonumber \\[2mm]}

\title{Improved Fixed Center Approximation of the Faddeev equations for the $\bar{K}NN$ system with $S=0$.}

\author{M. Bayar$^1$,$^2$ and E. Oset$^1$}
\affiliation{$^1$Instituto de F{\'\i}sica Corpuscular (centro mixto CSIC-UV)\\
Institutos de Investigaci\'on de Paterna, Aptdo. 22085, 46071, Valencia, Spain,\\ $^2$Department of Physics, Kocaeli University, 41380 Izmit, Turkey}

\begin{abstract}
We extend the Fixed Center Approximation (FCA) to the Faddeev Equations for the $\bar{K} NN$ system with $S=0$, including the charge exchange mechanisms in the $\bar{K}$ rescattering which have been ignored in former works within the FCA. We obtain similar results to those found before, but the binding is reduced by 6 MeV. At the same time we also evaluate the explicit contribution the $\pi N \Sigma$ intermediate state in the three body system and find that it produces and additional small decrease in the binding of about 3 MeV. The system appears bound by about 35 MeV and the width omitting two body absorption, is about 50 MeV.   

\end{abstract}


\vspace{1cm}

\date{\today}

\maketitle

\section{Introduction}
 \label{sec:intro}

  In this work we present a derivation of the Fixed Center Approximation to the Faddeev equations for the $\bar{K} NN$ system for the case when the $NN$ system has total spin $S=0$. The case for $S=1$ was done in \cite{kamalov} for $K^- d$ at threshold and extended to values below threshold to find a quasibound state in \cite{trento}. An earlier version for $S=0,1$ was done in \cite{melahat} but ignored charge exchange processes in the multiple scattering of the kaons. The improved results of \cite{trento} for $ S=1 $ provided a binding of about 12 MeV smaller than ignoring the charge exchange processes and was worth being taken into consideration. The resulting binding was 9 MeV with a width of 30 MeV. Yet, a formula like the one of \cite{kamalov} for the case of $S=0$ was never derived and one is left with some worry about what could be the effects of charge exchange processes in this latter case. This is the purpose of the present paper, where a simple derivation is presented and the new results are shown. 
  
 The study of the bound $\bar{K} NN$ system has been the center of much attention recently  \cite{Ikeda:2007nz,Shevchenko:2006xy,Shevchenko:2007zz,Dote:2008in,Dote:2008hw,Ikeda:2008ub,Ikeda:2010tk,newshev}. Earlier calculations on this system were already done in the 60's 
\cite{nogami} and more recently in \cite{akayama}, yet, the works of \cite{Ikeda:2007nz,Shevchenko:2006xy,Shevchenko:2007zz,Dote:2008in,Dote:2008hw,Ikeda:2008ub,Ikeda:2010tk} improved considerably on these earlier works.
    Two different methods have been used in those papers. Those of Refs.  
\cite{Ikeda:2007nz,Shevchenko:2006xy,Shevchenko:2007zz,Ikeda:2008ub} use Faddeev equations, in the formulation of Alt-Grassberger-Sandhas 
(AGS) \cite{Alt:1967fx}, using a separable interaction for the potentials with energy independent strength, and form factors depending on 
the three momenta. They include as coupled channels $ N \pi \Sigma$ and $NN \bar{K}$, the basic ones in the study of the $\bar K N$ interaction in chiral unitary theories at low energies \cite{Kaiser:1995eg,angels,ollerulf,Lutz:2001yb,Oset:2001cn,Hyodo:2002pk,Jido:2003cb,Borasoy:2005ie,Oller:2006jw,Borasoy:2006sr,ikedahyodo}. On the other hand the works of \cite{Dote:2008in,Dote:2008hw} use a variational method to obtain the binding energy and width, by means of an effective potential
 \cite{Hyodo:2007jq} that leads to the strongly energy dependent $\bar{K}N$ amplitudes of the chiral unitary approach.
 This includes
 the $\pi \Sigma$ channel integrated into the effective $\bar{K}N$ potential, but do not include the $N \pi \Sigma$ channel explicitly
 into the three body system. In Table IV of \cite{Shevchenko:2007zz} one finds that there are 11 MeV difference considering
just one channel $\bar{K}N$ or the two channels $\bar{K}N$ and $\pi \Sigma$, both in the evaluation of the $\bar{K}N$ amplitude and in the 
$\bar{K} NN-\pi \Sigma N$ system, finding more binding in the case of two channels. One should note that this is not to be compared with the 
 approach of \cite{Dote:2008in,Dote:2008hw} where the $\bar{K}N$ amplitude is always calculated with $\bar{K}N$, $\pi \Sigma$
and other coupled channels. One may argue that the consideration of $\pi \Sigma$ in the two body system would account for most of the $\pi \Sigma$ influence in the three body problem, a conjecture made by Schick and Gibson \cite{Schick:1978wi} which was
substantiated numerically in the work of Ref. \cite{Toker:1981zh}. But this was only done at threshold of $\bar K N$, so testing this at lower energies is worthwhile and we shall face this problem here too.

 The two Faddeev approaches \cite{Ikeda:2007nz,Shevchenko:2006xy} lead to binding energies higher than the variational approach, 50-70 MeV versus around 20 MeV binding
 respectively. The use of an energy independent kernel in the AGS equations is partly responsible for the extra
 binding of these approaches with respect to the chiral calculations.  
 The chiral potential is energy dependent, proportional to the sum of the two external meson energies in $\bar{K}N \to \bar{K}N$, and,
 as a consequence, one obtains a smaller  $\bar{K}N \to \bar{K}N$ amplitude at  $\bar{K} N$ energies below threshold, resulting in a smaller binding 
for the $\bar{K}NN$ system \cite{Dote:2008hw}. This is also corroborated in 
the approach of \cite{Ikeda:2007nz,Ikeda:2008ub} when the energy dependence of the Weinberg-Tomozawa chiral potential is taken into account
 in \cite{Ikeda:2010tk}. Further discussions along this issue can be found in section 5 of \cite{trento}.

Yet, in most cases the widths are 
systematically larger than the binding energy, of the order of 70-100 MeV, which makes the observation of these states problematic. A discussion on  experimental work claiming to have found a $\bar K NN$ bound state is made in section 5 of \cite{trento}, together with the theoretical work that shows that the experimental peaks observed correspond to conventional mechanisms that have nothing to do with a new bound state.

   The approach of \cite{melahat} follows the chiral unitary approach for the $\bar{K}N $ amplitudes, which provide the most 
important source of the binding of the three body system.  However, rather than   looking for the binding,
 searching for poles in the complex plane or looking for the energy that minimizes the expectation value of the Hamiltonian as the other approaches do,  one looks for peaks in the scattering matrices as a function of
 the energy of the three body system. These amplitudes  could in principle be used as input for final state 
interaction when evaluating cross sections in reactions where eventually this  $\bar{K}N N$ state is formed.

\section{Fixed Center formalism for the $\bar{K} (N N)$ system with $S=0$}

  We rely upon the results of 
\cite{MartinezTorres:2007sr,Khemchandani:2008rk,MartinezTorres:2008gy} which prove that only on shell two body amplitudes are needed to study the three body system. By this we mean the part of the analytical amplitudes obtained setting $q^2=m^2$ for the external particles, even if the particles are below threshold.

 Like in the other works, we also assume that the two body interactions proceed in L=0. 
  According to all the works, the main component of the wave function corresponds to having a $\bar{K} N $ in I=0 (coupling strongly to the $\Lambda(1405)$ and hence the total isospin will be I=1/2. The spin of the system for L=0 is provided by the two nucleons and all the works find that the most bound case corresponds to $S=0$. Since the $S=1$ case was studied in \cite{trento}, here we concentrate in the $S=0$ case. Since in our case $S_{NN}$ corresponds to the total spin of the $\bar{K}NN$ system we shall refer to $ S $ in what follows.

   In the FCA the $\bar{K}NN$ system is addressed by studying  the interaction of a $\bar{K}$ with a $NN$ cluster.
 The scattering matrix for this system is evaluated as a function of the total energy of the  $\bar{K}NN$ system and one looks for peaks in $|T|^2$.
In a second step one allows explicitly the intermediate  $\pi\Sigma N$ state in the three body system.
 A discussion of different approaches along the FCA line is done in \cite{Gal:2006cw}, where it is shown that some works 
do not take into account explicitly the charge exchange $K^{-}p\rightarrow\bar{K}^{0}n$ reactions and antisymmetry of the nucleons, but
it is explicitly done in Ref. \cite{kamalov}. In \cite{melahat} charge exchange in the  $\bar{K}$ multiple scattering was also ignored, but in \cite{trento} it was taken into account for the $S=1$ case, and a quasibound state was found with about 9 MeV binding and a width of about 30 MeV.

Technically we follow closely the formalism of \cite{multirho}, where the FCA has been considered, using chiral amplitudes, in order to
 study theoretically the possibility of forming multi-$\rho$ states with large spin.

The $NN$ interaction is of long range and very strong.
It binds the deuteron in spin $S=1$ and $I=0$, with $L=0$ to a very good approximation.
It nearly binds the $pp$ system also.
The binding of the $pp$ system is so close that a little help from an extra interaction, the strongly attractive ${\bar K}N$ interaction, is enough to also bind this system and we shall assume that this is the case here for $pp$ or in general for two nucleons in $S=0$, $I=1$, $L=0$.

The strategy followed in \cite{melahat} is to assume as a starting point that the $NN$ cluster has a wave function like the one of the deuteron (we  omit the d-wave).
Later on one releases this assumption and assumes that the $NN$ system is further compressed in the ${\bar K}NN$ system.
For this we rely upon the information on the $NN$ radius in the ${\bar K}NN$ molecule obtained in \cite{Dote:2008hw}, where the $NN$
interaction is taken into account including short range repulsion that precludes unreasonable compression. The r.m.s radius found is of 
the order of 2.2 fm, slightly above one half the value of the deuteron r.m.s radius of 3.98 fm ($NN$ distance).

\section{Calculation of the three body amplitude}

In order to evaluate the amplitude for the $\bar{K}NN$ with $S=0$, we follow the following strategy. 

1) We evaluate first the $K^- p p \rightarrow K^- p p $ amplitude considering charge exchange processes in the rescattering of the kaons. The amplitude will contains total isospin $I=1/2$ and $I=3/2$. Since we only want the $I=1/2$, we evaluate the scattering amplitude for $I=3/2$ in addition, taking the $\bar{K}^0 p p \rightarrow \bar{K}^0 p p $ amplitude, and from a linear combination of the two we obtain the $I=1/2$ amplitude. We have found that this strategy leads to a far simpler derivation than other methods that we have also tried. 

2) For the $K^- p p \rightarrow K^- p p $ we define three partition functions, 

a) $T_p$, which contains all diagrams that begin with a $K^-$ collision  with the first proton of the $pp$ system and finish with $K^- p p$.

b) $T_{ex}^{(p)}$, which contains all the diagrams that begin with a $\bar{K}^0 p$ collision on a $n p$ system and finish with $K^- p p$.

c) $T_{ex}^{(n)}$, which contains all the diagrams that begin with a $\bar{K}^0$ collision on a neutron  and finish with 
$K^- p p $.

\begin{figure}
\centering
\includegraphics[width=1.0\textwidth] {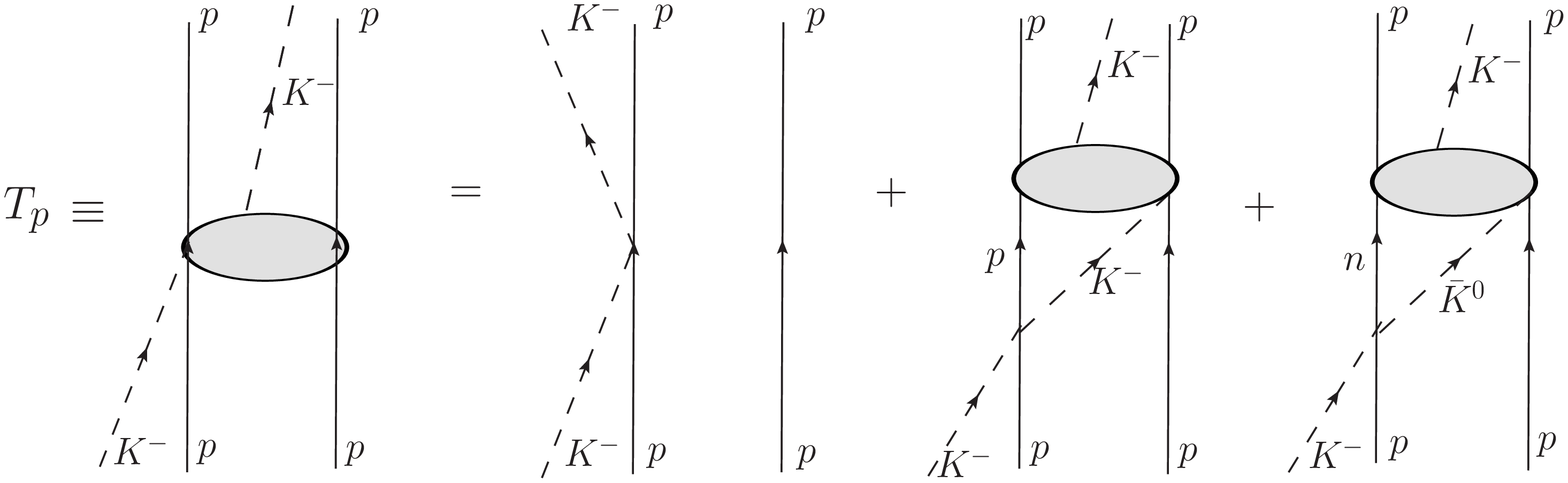}\\
\includegraphics[width=0.6\textwidth] {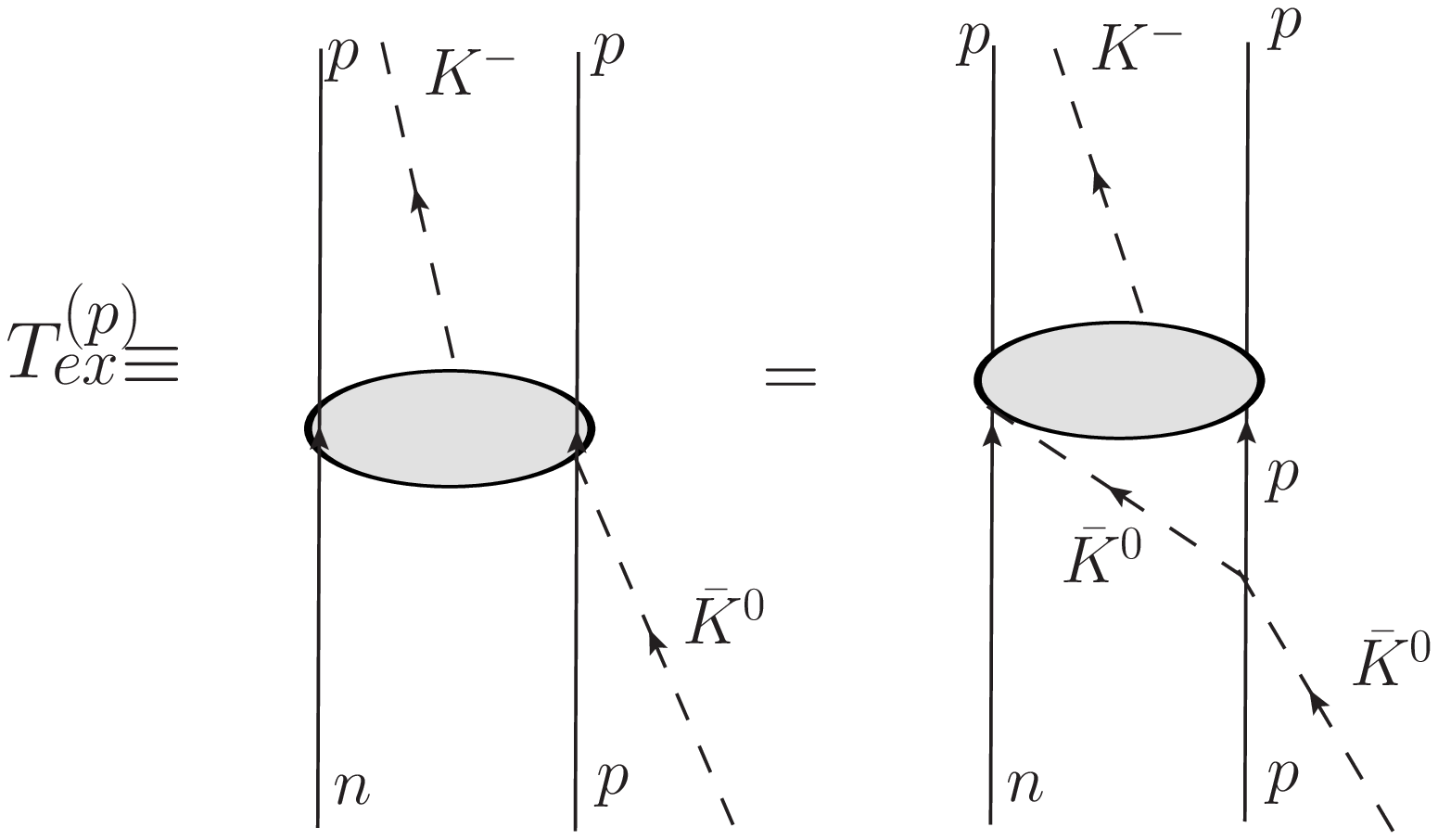}\\
\includegraphics[width=1.0\textwidth] {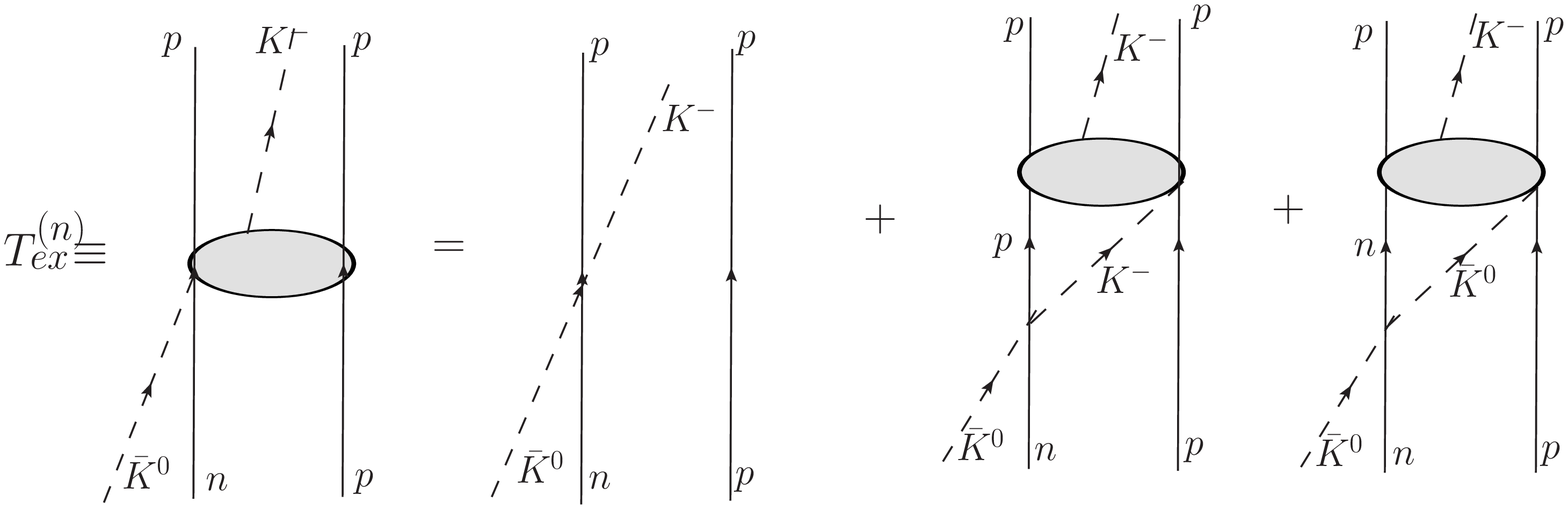}
\caption{Diagrammatic representation of the partition functions for the $K^- p p \rightarrow K^- p p $.}
\label{fig:tpp22}
\end{figure}

These amplitudes fulfill a set of coupled equations that are easy to see diagrammatically by looking at the diagrams of Fig. \ref{fig:tpp22}.

We have: 

\begin{eqnarray}
T_{p}&=&t_{p}+t_{p}G_0T_{p}+t_{ex}G_0T_{ex}^{(p)}\nonumber\\
T_{ex}^{(p)}&=&t_{0}^{(p)}G_0T_{ex}^{(n)}\nonumber\\
T_{ex}^{(n)}&=&t_{ex}+t_{ex}G_0T_{p}+t_{0}^{(n)}G_0T_{ex}^{(p)}
\label{Eq:tptexp}
\end{eqnarray} 
where
$t_p=t_{K^- p , K^- p}$, $t_{ex}=t_{K^- p , \bar{K}^0 n}$, $t_{0}^{(p)}=t_{\bar{K}^0 p , \bar{K}^0 p}$, $t_{0}^{(n)}=t_{\bar{K}^0 n , \bar{K}^0 n}$
and $G_0$ is the function \cite{multirho,melahat}

\begin{equation}
G_0=\int\frac{d^3q}{(2\pi)^3}F_{NN}(q)\frac{1}{{q^0}^2-\vec{q}\,^2-m_{\bar K}^2+i\epsilon}.
\label{Eq:gzero}
\end{equation}
where $F_{NN}(q)$ is the form factor of the $NN$ system. We can see that Eq. (\ref{Eq:gzero}) contains the folding of the $\bar{K}$ intermediate propagator with the form factor of the $NN$ system. The value of the 
$q^0$ in the total rest frame is given by 
\begin{equation}
q^0=\frac{s+m_{\bar K}^2-(2M_N)^2}{2\sqrt{s}}. 
\label{Eq:q00}
\end{equation}
The argument of the $  t^{(0)}$, $ t^{(1)} $ amplitudes is obtained following the steps of \cite{melahat, multirho} and is given by 

\begin{equation}
s_1=m^2_{\bar K}+M_N^2+\frac{1}{2}(s-m_{\bar K}^2-4M_N^2)~~.
\label{Eq:argment}
\end{equation}

By taking into account that $|K^->=-|1/2,~1/2>$ in the isospin basis, we can write all the former elementary amplitudes in terms of $I=0,~1$ ($t^{(0)},~t^{(1)}$) for the $\bar{K}N$ system, and we find

\begin{eqnarray}
t_{p}&=&\frac{1}{2}(t^{(0)}+t^{(1)})\nonumber\\
t_{ex}&=&\frac{1}{2}(t^{(0)}-t^{(1)})\nonumber\\
t_{0}^{(p)}&=&t^{(1)}\nonumber\\
t_{0}^{(n)}&=&\frac{1}{2}(t^{(0)}+t^{(1)}).
\label{Eq:tmatpex}
\end{eqnarray}

Solving Eq. (\ref{Eq:tptexp}) we find 

\begin{equation}
T_{p}=\frac{t_{p}(1-t_{0}^{(n)}G_0t_{0}^{(p)}G_0)+t_{ex}^2G_0t_{0}^{(p)}G_0}{(1-t_{p}G_0)(1-t_{0}^{(n)}G_0t_{0}^{(p)}G_0)-t_{ex}^2t_{0}^{(p)}G_0^3}
\label{Eq:tpp}
\end{equation}
which in isospin basis can be simplified to

\begin{equation}
T_{p}=\frac{\frac{1}{2}(t^{(0)}+t^{(1)})-t^{(0)}t^{(1)^2}G_0^2}{(1-G_0t^{(1)})(1+\frac{1}{2}(t^{(1)}-t^{(0)})G_0-G_0^2t^{(0)}t^{(1)})}
\label{Eq:tppisosipin}
\end{equation}

The total $K^- p p \rightarrow K^- p p $ amplitude would be $2T_{p}$ accounting for the first interaction of the $K^-$ with either of the protons.

3) Now we take into account that in the basis of $|I_{tot},I_{3,tot}>$
\begin{equation}
|K^- p p>=-(\dfrac{1}{\sqrt{3}}|3/2,1/2>+\sqrt{\dfrac{2}{3}}|1/2,1/2>)
\end{equation}
and thus
 \begin{equation}
<1/2|T|1/2>=\dfrac{3}{2}(<K^- p p|T|K^- p p> -\dfrac{1}{3}<3/2|T|3/2>)
\label{Eq:t1b2}
\end{equation}

\begin{figure}
\centering
\includegraphics[scale=0.6]{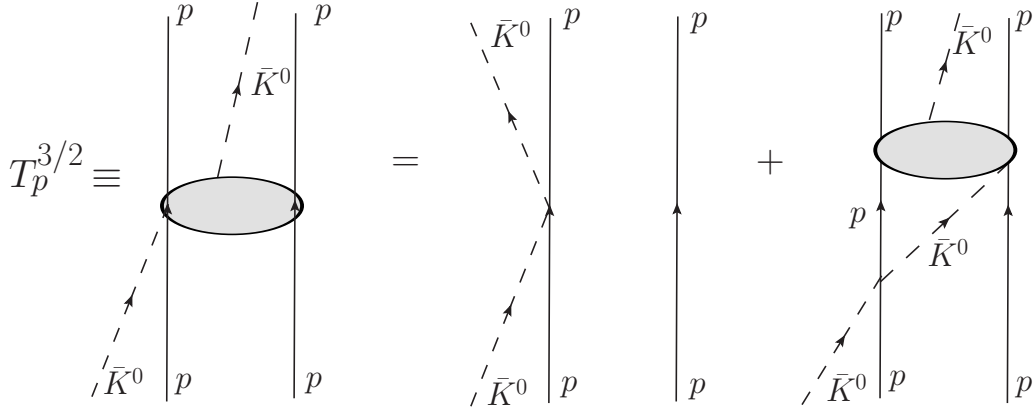}
\caption{Diagrammatic representation of the partition function for I=3/2.}\label{tpp3b2}
\end{figure}

The $<3/2|T|3/2>$ amplitude is particularly easy to obtain. In this case we take the $\bar{K}^0 p p \rightarrow \bar{K}^0 p p $ transition and diagrammatically we have the mechanism of Fig. \ref{tpp3b2} for the only partition function $T_p^{(3/2)}$.

Hence
\begin{equation}
T_p^{(3/2)}=t_{0}^{(p)}+t_{0}^{(p)}G_0T_p^{(3/2)}
\end{equation}
and the total $T^{(3/2)}$ amplitude will be  $2T_p^{(3/2)}$, accounting for the $\bar{K}^0$ interacting first also with the second nucleon. We have 

\begin{equation}
T_p^{(3/2)}=\dfrac{t_{0}^{(p)}}{1-G_0t_{0}^{(p)}}=\dfrac{t^{(1)}}{1-G_0t^{(1)}}.
\end{equation}
We can now use Eq. (\ref {Eq:t1b2}) and find for the total amplitude (including the factor two for first interaction with either proton)

\begin{eqnarray}
T^{(1/2)}&=&3T_p-\dfrac{t^{(1)}}{1-G_0t^{(1)}}\nonumber\\
&=&\dfrac{\dfrac{3}{2}t^{(0)}+\dfrac{1}{2}t^{(1)}-\dfrac{1}{2}t^{(1)}(t^{(1)}-t^{(0)})G_0-2t^{(0)}t^{(1)^2}G_0^2}{(1-G_0t^{(1)})(1+\frac{1}{2}(t^{(1)}-t^{(0)})G_0-G_0^2t^{(0)}t^{(1)})}
\end{eqnarray}
which can be simplified dividing the numerator by $(1-G_0t^{(1)})$ with the final result

\begin{eqnarray}
T^{(1/2)}&=&\dfrac{\dfrac{3}{2}t^{(0)}+\dfrac{1}{2}t^{(1)}+2G_0t^{(0)}t^{(1)}}{1+\frac{1}{2}(t^{(1)}-t^{(0)})G_0-G_0^2t^{(0)}t^{(1)}}.
\label{Eq:tknncharge}
\end{eqnarray}
This equation improves over the one obtained in \cite{melahat} although they coincide for first scattering term.

\section{Explicit consideration of the $\pi \Sigma N $ channel in the three body system}

\begin{figure}
\centering
\includegraphics[width=0.25\textwidth]{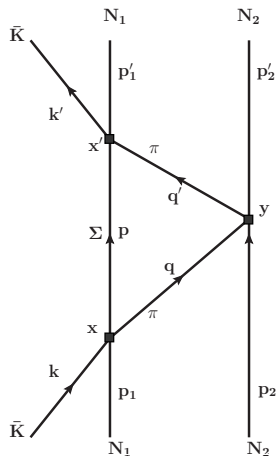}
\caption{Feynman diagram  for the $\pi \Sigma N$  channel.
The equivalent diagrams where the ${\bar K}$ interacts first with the second nucleon should be added.}
\label{fig:10}
\end{figure}

The $\pi \Sigma$ channel and other coupled channels are explicitly taken into account in the consideration of the 
$\bar{K}N$ amplitude which we have used in the FCA approach. This means that we account for the $\bar{K}N \rightarrow  \pi \Sigma$ transition,
and an intermediate  $\pi \Sigma N $ channel, but this  $\pi \Sigma $ state is again reconverted to $\bar{K}N$ leaving the other $N$
as a spectator. This is accounted for in the multiple scattering in the  $\bar{K}N \rightarrow \bar{K}N $ scattering matrix on one nucleon. However, we do not consider the possibility
that one has $\bar{K}N \rightarrow  \pi \Sigma$ and the $\pi$ rescatters with the second nucleon. If we want to have a final $\bar{K}N N$
system again, the $\pi$ has to scatter later with the $\Sigma$ to produce $\bar{K}N$. One may consider multiple scatterings of the $\pi$ 
with the nucleons, but given the smallness of the $\pi N$  amplitude compared to the $\bar{K}N$, any diagram  having more than one rescattering
of the pion with the nucleon will be negligible. This means that to account for explicit $\pi \Sigma N$ state in the three body system it is sufficient to consider the diagram  of
Fig. ~\ref{fig:10} (for $\bar{K}N$ scattering on nucleon 1). 

The explicit expression for this contribution can be seen \cite{melahat} (section IV). In particular it involves the isospin combination in the $\bar{K}N,\pi \Sigma $ 

\begin{equation}
t_{\bar{K}N,\pi \Sigma}=\dfrac{1}{4}t_{\bar{K}N,\pi \Sigma}^{I=1}+\dfrac{3}{4}t_{\bar{K}N,\pi \Sigma}^{I=0}~;~S=0
\label{Eq:isospins0}
\end{equation}
We note, in passing, that the combination found for $ S=1 $ was 
\begin{equation}
t_{\bar{K}N,\pi \Sigma}=\dfrac{3}{4}t_{\bar{K}N,\pi \Sigma}^{I=1}+\dfrac{1}{4}t_{\bar{K}N,\pi \Sigma}^{I=0}~;~S=1
\label{Eq:isospins1}
\end{equation}
and the $ \pi N,\pi N $ amplitudes were also different. This justifies the different effects found for these two cases, although the changes are small in both of them. Non-perturbatively we can add   $2 T^{\pi \Sigma}$ of \cite{melahat} to Eq. (\ref{Eq:isospins0}). In as much as the term $ \frac{3}{2}t^{(0)}$ is the largest component of Eq. (\ref{Eq:tknncharge}), if we add  $2 T^{\pi \Sigma}$ to $ \frac{3}{2}t^{(0)}$ in all terms in  Eq. (\ref{Eq:tknncharge}), we guarantee the correct results for the lowest order contribution and account approximately for multiple scattering including the $T^{\pi \Sigma}$  contribution. We have seen that the perturbative and non-perturbative results are very similar, but there is a difference of about 2 MeV in the position of the peak (the non-perturbative case more bound).

\section{Results for the $\bar{K}NN$ system with $S=0$}

 \begin{figure}
\centering
\includegraphics[width=1\textwidth]{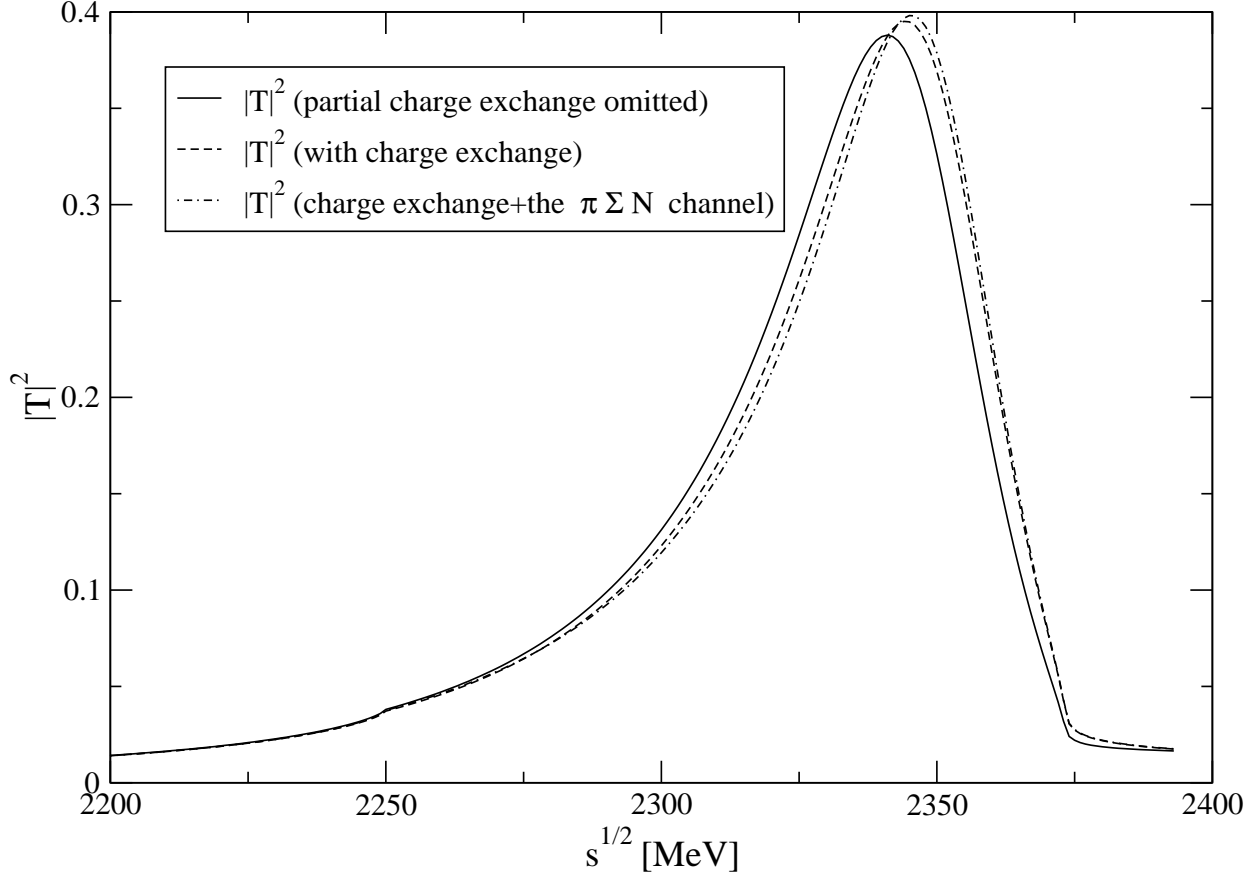}
\caption{Modulus squared of the three-body scattering amplitude for singlet (S=0) with deuteron size for $NN$.  The solid line indicates the $\bar{K}NN$ system  with partial omission of  charge exchange \cite{melahat}, the dashed line indicates the same system taking into account charge exchange diagrams  and the dot-dashed line indicates the result of the including
the  $\pi \Sigma N$ channel and charge exchange diagrams.}
\label{fig:s0}
\end{figure} 

\begin{figure}
\centering
\includegraphics[width=1\textwidth]{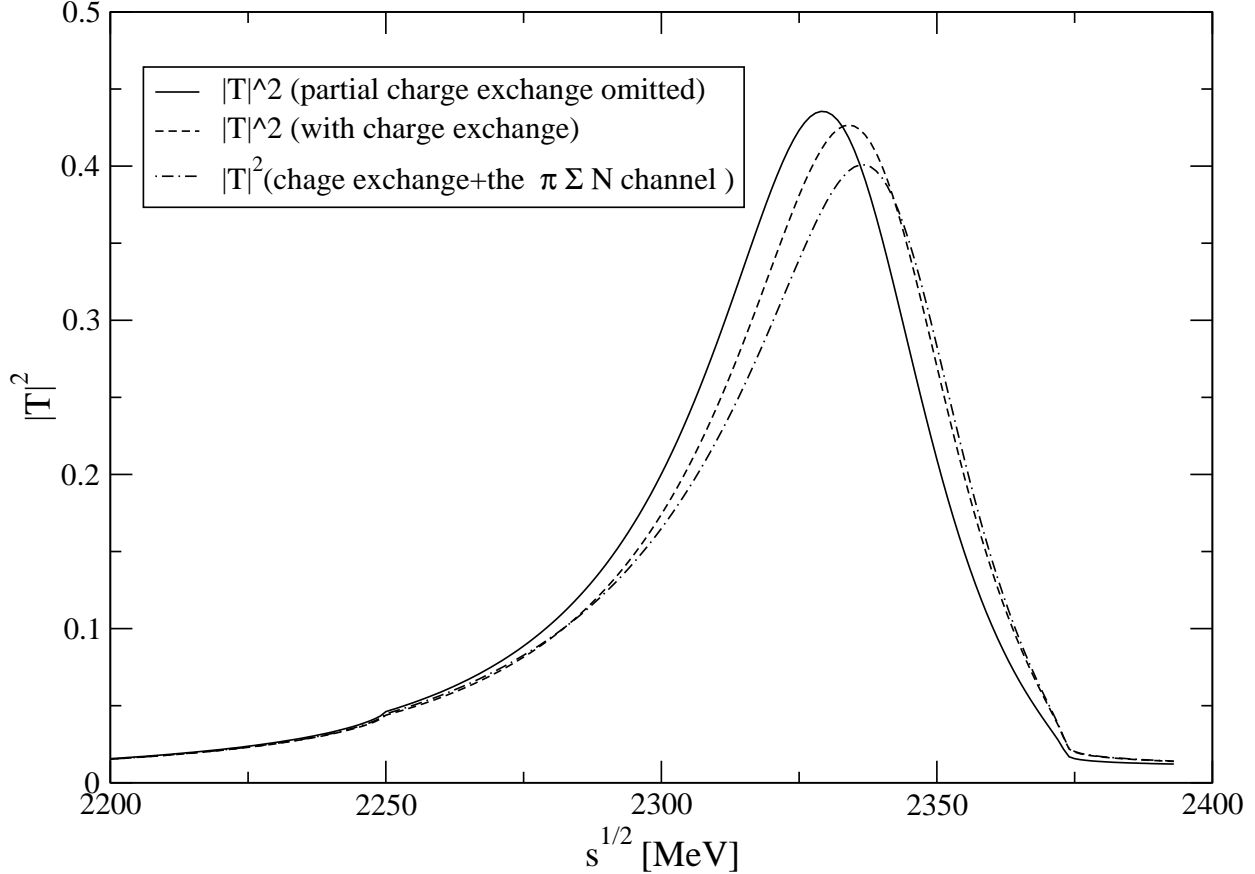}
\caption{Modulus squared of the three-body scattering amplitude for singlet (S=0) with reduced $NN$ size from \cite{Dote:2008hw}.  The solid line indicates the $\bar{K}NN$ system with partial omission of  charge exchange , the dashed line indicates the same system taking into account charge exchange diagram and the dot-dashed line indicates the result of the including
the  $\pi \Sigma N$ channel and charge exchange diagrams.}
\label{fig:s0red}
\end{figure}

In Figs. \ref{fig:s0} and  \ref{fig:s0red} we show our results. In Fig. \ref{fig:s0} we use the form factor of the deuteron for the $NN$ system and we show the results for $|T|^2$ in three options: a) Partial omission of charge exchange, using the expression obtained in \cite{melahat}

\begin{equation}
T^{(pceo)}=\frac{2 t}{1-G_0t}~;~~~t=\frac{1}{4} t^{(1)}+\frac{3}{4} t^{(0)}
\label{Eq:tknneski}
\end{equation}
By partial omission we mean that some charge exchange is accounted for when making the transition in the single 
scattering from $ K^- pp $ to $\bar{K}^0 pn$, the two components of the $ \bar{K}(NN(I=1)) $ wave function. However, charge exchange in double (in multiple) scattering in the diagonal transitions $ K^- pp \rightarrow K^- pp $ would be neglected. This is unlike the $ \bar{K}(NN(I=0)) $ system where the  $ K^- (pn-np) (I=0) \rightarrow \bar{K}^0 nn$ transition in single scattering is not allowed;
b) charge exchange explicitly included, Eq. (\ref{Eq:tknncharge});
c) explicit $\pi \Sigma N$ channel added to the amplitude of Eq. (\ref{Eq:tknncharge}), as we have described in the former section.

We can see from the figure that the consideration of charge exchange processes reduces the binding in 4 MeV with respect to the results in \cite{melahat}. This is smaller than the 12 MeV shift found in \cite{trento} for the $S=1$ case. Also, relatively, the change is more drastic in the $S=1$ case since there was a reduction of 12 MeV  from 21 MeV, while here there is a reduction of 4 MeV over a total of 32 MeV binding. When we consider the explicit intermediate $\pi \Sigma N$ state we find an extra reduction of 1 MeV in the binding. 

The results are similar, but the differences more pronounced, when we redo the calculations using the reduced $NN$ radius of \cite{Dote:2008hw} (see \cite {melahat} for details to implement the form factor with the reduced radius). As we can see in Fig. \ref{fig:s0red}, the binding goes now from about 44 MeV when partial omission of charge exchange is considered, Eq. (\ref{Eq:tknneski}), to 38 MeV when it is explicitly considered, and then to 35 MeV  when the $\pi \Sigma N$ channel is explicitly taken into account. The latter case is our final result, which provides 35 MeV binding for this system and a width of about 50 MeV. 

\begin{figure}
\centering
\includegraphics[width=1\textwidth]{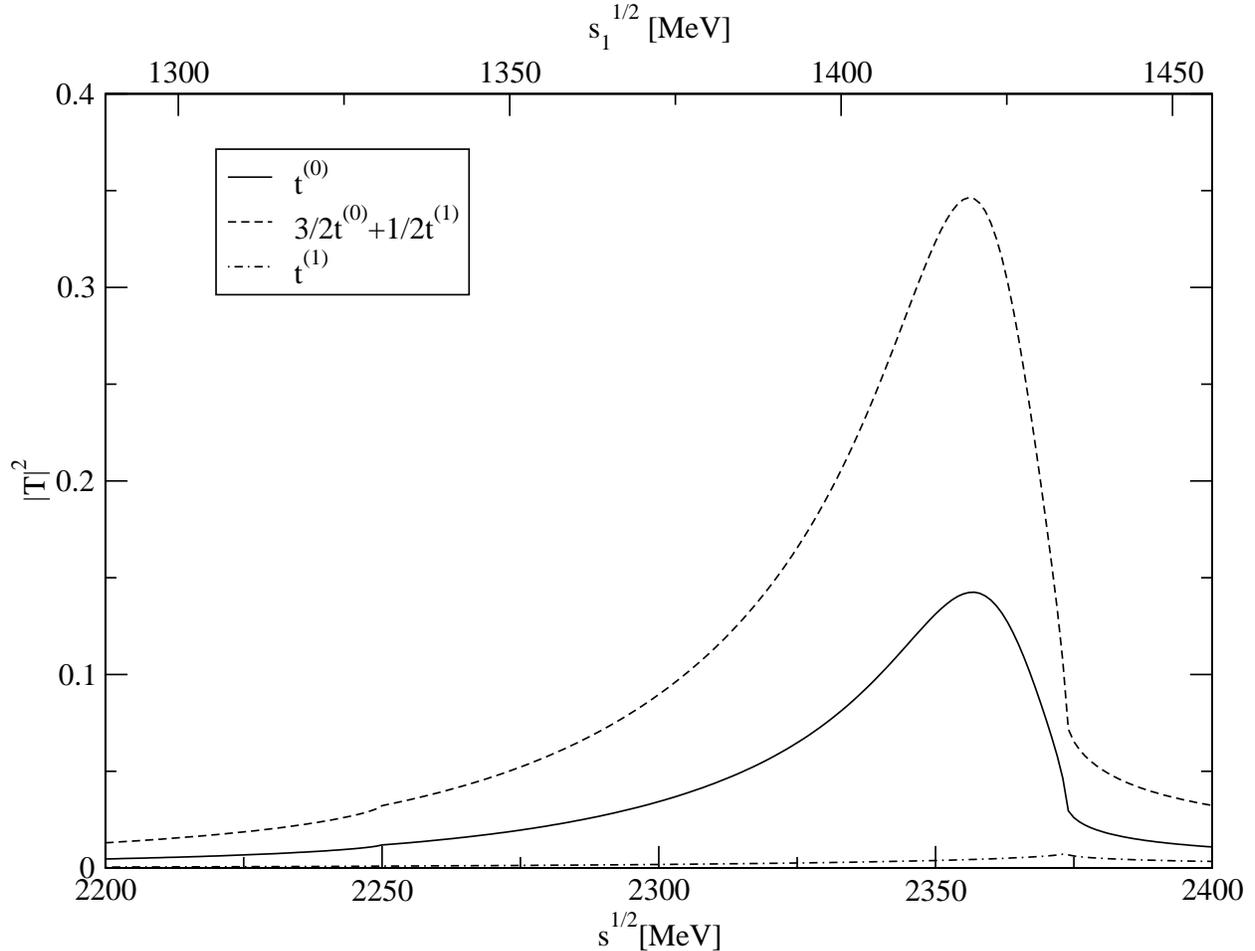}
\caption{Single scattering amplitude: Modulus squared of $ t^{(0)} $, $ t^{(1)} $, and $ \frac{3}{2}t^{(0)}+ \frac{1}{2}t^{(1)}$. The argument is $ \sqrt{s} $ for the three body system. The argument $ \sqrt{s_{1}} $ for the two body system is  obtained through Eq. (\ref{Eq:argment}). The upper x-axis scale is for  $\sqrt{s_{1}} $.}
\label{fig:kbarn}
\end{figure} 

It is interesting to see the origin of the peaks in Figs. \ref{fig:s0}, \ref{fig:s0red}. For this purpose we show in Fig. \ref{fig:kbarn} the modulus squared of the $ t^{(0)} $, $ t^{(1)} $, and $ \frac{3}{2}t^{(0)}+ \frac{1}{2}t^{(1)}$  amplitudes. We see that the shape of  Figs. \ref{fig:s0}, \ref{fig:s0red} is similar to the one of the $ t^{(0)} $, which already provides a peak in the total amplitude of Eq. (\ref{Eq:tknncharge}) at the single scattering level. The multiple scattering is responsible for an extra binding of about 25 MeV and 10 MeV increase in the width.
Alternatively, we can view the $\bar{K}NN$ state as a bound $ \Lambda(1405) N $ state as suggested in \cite{oka1, oka2}. In this case Fig. \ref{fig:kbarn} could be interpreted as having a non interacting $ \Lambda(1405) N $  state, where the $ \Lambda(1405)$ has been formed in the collision of the $\bar{K}$ with one nucleon and the second one has been a spectator. Then, the results of Fig. \ref{fig:s0red}, where the $\bar{K}$ has reinteracted with the two nucleons, could be interpreted as a system where the  $ \Lambda(1405)$ has interacted with the second nucleon and a quasibound $ \Lambda(1405) N $ state is formed \cite{oka2}.

\section{Discussion and Conclusions} 

We have done a derivation of the FCA to the Faddeev equations for the case of the $\bar{K}NN$ system with $S=0$, complementing what was already done in \cite{kamalov} for the $K^- d$ at threshold, extended below threshold in \cite{trento}. With this new formula we evaluate the $\bar{K}NN$ amplitude below the $\bar{K}NN$ threshold and we find a clear peak of $|T|^2$ at about  3334 MeV, reducing the former binding by about 6 MeV,  with a width of 50 MeV.  This is a moderate change with respect to the one found in \cite{melahat} ignoring the charge exchange processes, and also smaller than the effect found in \cite{trento} for the $S=1$, where the charge exchange processes was found quite important, reducing the binding of the $\bar{K}NN$ with $S=1$ in about 12 MeV.
  We have also shown that the explicit consideration of the $\pi N \Sigma$ state in the three body system is rather moderate, and goes in the direction of decreasing the binding by about  3 MeV.
  In summary, here and in \cite{trento} we have shown that the FCA to the Faddeev equations is a good tool which provides reasonable values for the binding and width of the  $\bar{K}NN$ system, corroborating qualitatively findings done with other methods which are technically much more involved. We find a binding for the $\bar{K}NN$ system with  $S=0$ of 35 MeV with a width of 50 MeV.  
 This should be compared to the about 20 MeV binding in \cite{Dote:2008in, Dote:2008hw} and 16 MeV in a recent paper \cite{Barnea:2012qa} using the same input as in \cite{Dote:2008in, Dote:2008hw}. Even with these differences,  the binding obtained here is substantially smaller than other values quoted in the literature which did not use the chiral dynamical approach. As to what to attribute the difference with the results of \cite{Dote:2008in, Dote:2008hw, Barnea:2012qa}, certainly there are approximations done in the FCA. One of them, making a static picture of the two $ NN $ and using Eq. (\ref{Eq:argment}), which relates the $\bar{K}N$ effective energy to the total energy, does not consider recoil of the nucleon.
 
  One can make some rough estimate of this effects as follows. Since $ s_{1}=(p_{\bar{K}}+p_{N_{1}})^{2} =(P-p_{N_{2}})^{2}$, with $ P $ the total fourmomentum of the $\bar{K}NN$ system ($ P^{2}=s $), we would have  
\begin{equation}
s_{1}=(\sqrt{s}-p_{N_{2}}^{0})^{2}-\vec p_{N_{2}}^{~2}
\label{Eq:yenis1}
\end{equation}
and the first term of Eq. (\ref{Eq:yenis1}) would correspond to the expression of Eq. (\ref{Eq:argment}) that neglects recoil. Continuing with the rough approximation we can think that 

 \begin{equation}
\dfrac{\vec p_{N_{2}}^{~2}}{2M_{N}}\simeq B_{N}\simeq M_{N}(\dfrac{2M_{N}+m_{\bar{K}}-\sqrt{s}}{2M_{N}+m_{\bar{K}}}),
\label{Eq:yenipn2}
\end{equation}
with $ B_{N} $ the binding energy of a single nucleon, 
where we have assumed the kinetic energy to be of the order of the binding energy and the binding energy of each particle proportional to its mass (this is also implicit in Eq. (\ref{Eq:argment}) (see \cite{melahat,multirho})).
 This correction results into a decrease of the binding of the $S=0$ state by about 8 MeV, bringing the results closer those of \cite{Dote:2008in, Dote:2008hw, Barnea:2012qa}. Incidentally, the $S=1$ state reported in \cite{trento} would reduce its binding by about 2-3 MeV. 
 
 On the other hand, the variational or Faddeev approaches also have intrinsic uncertainties tied to the use of off-shell extrapolation of the amplitudes.  
  To our knowledge there is only one work where the effects of these off-shell extrapolations has been studied in the three body system. This is the work of \cite{MartinezTorres:2008gy} (section V), which studies the $K \bar{K} \phi$ system with Faddeev equations, and the off-shell effects were shown to produce a shift of 40 MeV in the energy of the $ X(2175) $
 resonance (now called  $ \phi(2170) $).
 One can make some guesses on what the effects could be in the present case. The amount of 40 MeV over the excess energy of the  $ \phi(2170) $ over threshold of $K \bar{K} \phi$, 170 MeV, is about 24 \%. Thus, the same proportion applied to the binding of the $\bar{K}NN$ system found by us gives 8 MeV. Alternatively, we can look at 
the differences between different approaches using the same input to give us an idea of the possible effects. Inspection of the results of \cite{Dote:2008in, Dote:2008hw,Ikeda:2010tk, Barnea:2012qa} tells us that the differences are of the order of 10 MeV. Thus, effects of the order of 10 MeV from unphysical off-shell contributions are likely in the $ S=0 $ case.
 On the other hand, the $S=1$ case of \cite{trento} could be a privileged one in this sense, since applying the same percentage as before to the 9 MeV binding found in \cite{trento}, we find likely  off-shell effects of the order of 2 MeV.
 The findings for the $S=1$ in \cite{trento}, with the uncertainties estimated here could be in agreement with the results for this case in \cite{Barnea:2012qa},  where the authors conclude that if there is a quasibound $ K^{-} d$ state it must be bound by less than about 11 MeV.
  
  Within the uncertainties of the method, the simplicity of the present approach allows one to get an insight on the problems, following the analytical expressions of the amplitudes. 
  Ultimately, improvements on all these works could be done within the approach of  
\cite{MartinezTorres:2007sr,Khemchandani:2008rk,MartinezTorres:2008gy}, which  one must encourage, which uses full Faddeev equations in coupled channels and relies upon the on-shell t-matrices alone.

  \section{Acknowledgments}
  We thank A. Gal and T. Hyodo for useful comments. This work is partly supported by DGICYT contract number
FIS2011-28853-C02-01, the Generalitat Valenciana in the program Prometeo, 2009/090. We acknowledge the support of the European Community-Research Infrastructure
Integrating Activity
Study of Strongly Interacting Matter (acronym HadronPhysics3, Grant Agreement
n. 283286)
under the Seventh Framework Programme of EU.



\begin{thebibliography}{999}


\bibitem{kamalov}
  S.~S.~Kamalov, E.~Oset and A.~Ramos,
  Nucl.\ Phys.\  A {\bf 690}, 494 (2001).

\bibitem{trento} 
  E.~Oset, D.~Jido, T.~Sekihara, A.~M.~Torres, K.~P.~Khemchandani, M.~Bayar and J.~Yamagata-Sekihara,
   Nucl.\ Phys.\  A {\bf 881}, 127 (2012).


\bibitem{melahat}
  M.~Bayar, J.~Yamagata-Sekihara, E.~Oset,
  Phys.\ Rev.\  {\bf C84}, 015209 (2011).

\bibitem{Ikeda:2007nz}
  Y.~Ikeda and T.~Sato,
  Phys.\ Rev.\  C {\bf 76}, 035203 (2007).
  
\bibitem{Shevchenko:2006xy}
  N.~V.~Shevchenko, A.~Gal and J.~Mares,
  Phys.\ Rev.\ Lett.\  {\bf 98}, 082301 (2007).
  
\bibitem{Shevchenko:2007zz}
  N.~V.~Shevchenko, A.~Gal, J.~Mares and J.~Revai,
  Phys.\ Rev.\  C {\bf 76}, 044004 (2007).

  
\bibitem{Dote:2008in}
  A.~Dote, T.~Hyodo and W.~Weise,
  Nucl.\ Phys.\  A {\bf 804}, 197 (2008).

\bibitem{Dote:2008hw}
  A.~Dote, T.~Hyodo and W.~Weise,
  Phys.\ Rev.\  C {\bf 79}, 014003 (2009).

\bibitem{Ikeda:2008ub}
  Y.~Ikeda and T.~Sato,
  Phys.\ Rev.\  C {\bf 79}, 035201 (2009).
        
	
\bibitem{Ikeda:2010tk}
  Y.~Ikeda, H.~Kamano and T.~Sato,
  Prog.\ Theor.\ Phys.\  {\bf 124}, 533 (2010).

	
\bibitem{newshev}  N.~V.~Shevchenko,
 Phys.\ Rev.\ C {\bf 85}, 034001 (2012).
	
\bibitem{nogami} Y. Nogami, Phys. Lett. 7, 288 (1963).



  
\bibitem{akayama}
  T.~Yamazaki and Y.~Akaishi,
  Phys.\ Lett.\  B {\bf 535} (2002) 70.
  
\bibitem{Alt:1967fx}
  E.~O.~Alt, P.~Grassberger and W.~Sandhas,
  Nucl.\ Phys.\  B {\bf 2} (1967) 167.


 
\bibitem{Kaiser:1995eg}
  N.~Kaiser, P.~B.~Siegel and W.~Weise,
  Nucl.\ Phys.\  A {\bf 594}, 325 (1995).

\bibitem{angels}
  E.~Oset and A.~Ramos,
  Nucl.\ Phys.\  A {\bf 635}, 99 (1998).

\bibitem{ollerulf}
  J.~A.~Oller and U.~G.~Meissner,
  Phys.\ Lett.\  B {\bf 500}, 263 (2001).
  
\bibitem{Jido:2003cb}
  D.~Jido, J.~A.~Oller, E.~Oset, A.~Ramos, U.~G.~Meissner,
  Nucl.\ Phys.\  {\bf A725}, 181-200 (2003).



  
  
    

  
\bibitem{Oset:2001cn}
  E.~Oset, A.~Ramos and C.~Bennhold,
  Phys.\ Lett.\  B {\bf 527}, 99 (2002)
  [Erratum-ibid.\  B {\bf 530}, 260 (2002)].
 
\bibitem{Hyodo:2002pk}
  T.~Hyodo, S.~I.~Nam, D.~Jido and A.~Hosaka,
  Phys.\ Rev.\  C {\bf 68}, 018201 (2003).
  


\bibitem{Borasoy:2005ie}
  B.~Borasoy, R.~Nissler and W.~Weise,
  Eur.\ Phys.\ J.\  A {\bf 25}, 79 (2005).
  
  
\bibitem{Oller:2006jw}
  J.~A.~Oller,
  Eur.\ Phys.\ J.\  A {\bf 28}, 63 (2006).



\bibitem{Borasoy:2006sr}
  B.~Borasoy, U.~G.~Meissner and R.~Nissler,
  Phys.\ Rev.\  C {\bf 74}, 055201 (2006).
  
\bibitem{ikedahyodo} 
  Y.~Ikeda, T.~Hyodo and W.~Weise,
    Nucl.\ Phys.\  A {\bf 881}, 98 (2012).

  
\bibitem{Lutz:2001yb}
  M.~F.~M.~Lutz and E.~E.~Kolomeitsev,
  Nucl.\ Phys.\  A {\bf 700}, 193 (2002).
  
\bibitem{Hyodo:2007jq}
  T.~Hyodo and W.~Weise,
  Phys.\ Rev.\  C {\bf 77}, 035204 (2008).
  
\bibitem{Schick:1978wi}
  L.~H.~Schick, B.~F.~Gibson,
  Z.\ Phys.\  {\bf A288}, 307 (1978).


\bibitem{Toker:1981zh}
  G.~Toker, A.~Gal, J.~M.~Eisenberg,
  Nucl.\ Phys.\  {\bf A362}, 405-430 (1981).
  
\bibitem{MartinezTorres:2007sr}
  A.~Martinez Torres, K.~P.~Khemchandani, E.~Oset,
  Phys.\ Rev.\  {\bf C77}, 042203 (2008).
 
\bibitem{Khemchandani:2008rk}
  K.~P.~Khemchandani, A.~Martinez Torres, E.~Oset,
  Eur.\ Phys.\ J.\  {\bf A37}, 233-243 (2008).
  

\bibitem{MartinezTorres:2008gy}
  A.~Martinez Torres, K.~P.~Khemchandani, L.~S.~Geng, M.~Napsuciale, E.~Oset,
  Phys.\ Rev.\  {\bf D78}, 074031 (2008).
  
\bibitem{Gal:2006cw}
  A.~Gal,
  Int.\ J.\ Mod.\ Phys.\  {\bf A22}, 226-233 (2007).

\bibitem{multirho}
  L.~Roca and E.~Oset,
  Phys.\ Rev.\  D {\bf 82}, 054013 (2010).

\bibitem{oka1}  A.~Arai, M.~Oka and S.~Yasui,
 Prog.\ Theor.\ Phys.\  {\bf 119}, 103 (2008).

\bibitem{oka2}  T.~Uchino, T.~Hyodo and M.~Oka,
 Nucl.\ Phys.\ A {\bf 868-869}, 53 (2011).
\bibitem{Barnea:2012qa}
  N.~Barnea, A.~Gal and E.~Z.~Liverts,
   Phys.\ Lett.\  B {\bf 712} (2012) 132-137.
    

\end{thebibliography}
\end{document}